\documentclass{elsart}

 \usepackage{graphicx}

\usepackage{amssymb}

\begin{document}

\begin{frontmatter}



\title{Theory of spin polarization in mesoscopic spin-orbit coupling systems}


\author{Shi-Dong Liang\corauthref{cor1}}
\corauth[cor1]{Corresponding author Email: stslsd@mail.sysu.edu.cn}
and Guang-Yao Huang
\address{State Key Laboratory of Optoelectronic Material and Technology, and \\
Guangdong Province Key Laboratory of Display Material and Technology, \\
School of Physics and Engineering,
Sun Yat-Sen University, Guangzhou, 510275, People's Republic of China
}

\begin{abstract}
We establish a general formalism of the bulk spin polarization (BSP) and the
current-based spin polarization (CSP) for mesoscopic
ferromagnetic and spin-orbit interaction (SOI) semiconducting systems. Based on this formalism,
we reveal the basic properties of BSP and CSP and their relationships.
The BSP describes the intrinsic spin polarized properties of devices.
The CSP depends on both intrinsic parameters of device and the incident current.
For the non-spin-polarized incident current with the in-phase spin-phase coherence,
CSP equals to BSP. We give analytically the BSP and CSP of several typical nano
device models, ferromagnetic nano wire, Rashba nano wire and rings. These results
provide basic physical behaviors of BSP and CSP and their relationships.
\end{abstract}

\begin{keyword}
spin polarization; Spin polarized transport in semiconductors
\PACS 72.10.Bg;	72.25.Dc
\end{keyword}
\end{frontmatter}

\section{Introduction}
\label{}
The spin degree freedom of electron promises great opportunities for a new
paradigm of nanodevices and opens some fundamental issues in condensed
matter physics. The spin current generation, spin orientation control and
manipulation become challenging problems for spintronics. \cite{Prinz} The
spin current can be induced by the spin-orbit interaction (SOI) in some
low-dimensional semiconductors \cite{Winkler}, such as Rashba effect due to
the structure inversion asymmetry\cite{Rashba} and Dresselhaus effect due to
the bulk inversion asymmetry. \cite{Dresselhaus} Especially the Rashba SOI can
be modulated electrically in some semiconductor materials such as
InAs/GaSb, AlSb/InAs, GaAs/GaAlAs heterostructures, \cite{Luo} which offers
great potential applications. One of key problems is how to measure the spin
polarization in semiconductors. Theoretically, there are two
definitions of spin polarization wildly used in the early study on the spin-dependent
electronic transport in ferromagnetic metals and heterostructures\cite{Schmidt}.
One is called the bulk spin polarization (BSP) defined by \cite{Schmidt}

\begin{equation}
P_{g}=\frac{g^{\uparrow }-g^{\downarrow }}{g^{\uparrow }+g^{\downarrow }}
\label{bulk}
\end{equation}%
where $g^{\uparrow (\downarrow )}$ is the spin-dependent conductance.
This definition of spin polarization reflects an intrinsic material property.
It depends on the electron population and mobility in two spin states.
The other spin polarization is defined based on the spin current,\cite{Schmidt}
\begin{equation}
P_{j}=\frac{j^{\uparrow }-j^{\downarrow }}{j^{\uparrow }+j^{\downarrow }}
\label{current}
\end{equation}%
where $j^{\uparrow (\downarrow )}$ is the spin-dependent current.
This definition of spin polarization measures the
asymmetry of spin-up and -down currents. It may be regarded as the
current-based spin polarization (CSP). In a homogeneous material
with an electric bias $P_{g}=P_{j}$ . However, when the conductances of
interfaces between different material mismatch, $P_{g}\neq P_{j}$.
\cite{Schmidt} Actually, there are still arguments on how to measure the
spin polarization of the spin-dependent transports through the interfaces of
different ferromagnetic metals \cite{Schmidt2}, especially through the
interface of ferromagnetic metals and nonmagnetic semiconductors.\cite%
{Schmidt} Hence, it should be careful for the cases of $P_{g}\neq P_{j}$.
The questions are what are physical conditions for $P_{g}=P_{j}$ and what is
difference between BSP and CSP.

In this paper, we will establish a formalism of spin polarization for
ferromagnetic and mesoscopic SOI systems and analytically give the basic
properties of BSP and CSP in ferromagnetic and SOI semiconductor systems.
We will find a general relationship between BSP and CSP for both of the systems.
In Sec. II, making use of the wavefunction matching method,
we obtain the general formalism of spin polarization in mesoscopic systems
and give the analytic expressions of BSP and CSP as well as their relationships.
In Sec. III, the general formalism of spin polarization is applied to some typical
models of nano devices. Finally, we will give discussion and conclusion in Sec. IV.
Some mathematical derivations will be given in the appendices.

\section{Formalism of spin polarization}
The elementary model of spintronics devices consists of three parts, the central device,
the left and right electrodes. The spin-dependent incident current from the left electrode goes through the device to the
right electrode. The basic physical issue is how to control the spin current and how
to measure the spin polarization of this system.
We may ask two basic questions: what is difference between the
BSP and CSP? How does the central part affect the spin-dependent incident current?
To reveal the basic physics of spin polarization, we consider only a single mode quantum contact with
electrodes at Fermi energy. The spin-dependent conductance is proportional to
the transmission coefficient, $g^{\sigma }=\frac{e^{2}}{h}|T^{\sigma}|^{2}$, where $T^{\sigma}$ is the
spin-$\sigma$ transmission coefficient, which reflects an intrinsic electronic transport property of the device.
Thus, the BSP can be written as
\begin{equation}
P_{g}=\frac{|T^{\uparrow }|^{2}-|T^{\downarrow }|^{2}}{|T^{\uparrow}|^{2}+|T^{\downarrow}|^{2}}
\label{BSP}
\end{equation}%
The spin-dependent current is defined by
$j_{\mu }^{\sigma }=e\mathrm{Re}(\psi _{\mu }^{\sigma }v_{\mu }^{\sigma}\psi _{\mu}^{\sigma })$,
where $\mu \equiv in,out$ and $c$ label the left, the right
and the central parts. The velocity operators are given by $v^{\sigma}=\frac{\partial H}{\partial p}$.
In general, the Hamiltonian of heterojunction is\cite{Fabian}
\begin{equation}
H=-\frac{\hbar^{2}\nabla^{2}}{2m^{*}_{\mu}}+V_{\mu}+\mathbf{\Delta}_{\mu}\cdot \mathbf{\sigma},
\end{equation}
We consider that the spin states are degenerate, $\Delta_{\ell(r)}=0$ in the electrodes for common metals.
The Bloch wave functions of electrons in general are\cite{Molenkamp,Splettstoesser}
\begin{equation}
\Psi_{in}(x)=\sum_{\sigma}(A^{\sigma }e^{ikx}+B^{\sigma }e^{-ikx})\chi _{in}^{\sigma }
\end{equation}
for incident (left lead) $x<0$,
\begin{equation}
\Psi_{c}(x)=\sum_{\sigma}
(u^{\sigma }e^{ik_{+}^{\sigma }x}+v^{\sigma }e^{ik_{-}^{\sigma }x})\chi_{c}^{\sigma }
\end{equation}
for device (central) $x\in (0,L)$, and
\begin{equation}
\Psi_{out}(x)=\sum_{\sigma }t^{\sigma }e^{ikx}\chi _{out}^{\sigma }
\label{wave}
\end{equation}
for outgoing (right lead) $x>L$, where $L$ is the size of the device. Thus, the spin-dependent current can be
expressed as $j_{in}^{\sigma }=k|A^{\sigma }|^{2}$ and $
j_{out}^{\sigma }=k|t^{\sigma}|^{2}$. We have
$t^{\sigma}=T^{\sigma}A^{\sigma}$. Consequently, the CSP becomes
\begin{equation}
P_{j,out}=\frac{|T^{\uparrow}|^{2}|A^{\uparrow}|^{2}-|T^{\downarrow}|^{2}|A^{\downarrow}|^{2}}
{|T^{\uparrow}|^{2}|A^{\uparrow}|^{2}+|T^{\downarrow}|^{2}|A^{\downarrow}|^{2}}
\label{CSPout}
\end{equation}%
\begin{equation}
P_{j,in}=\frac{|A^{\uparrow}|^{2}-|A^{\downarrow}|^{2}}{|A^{\uparrow}|^{2}+|A^{\downarrow}|^{2}}
\label{CSPin}
\end{equation}%
The CSP depends on the amplitude of the incident current. In general,
the BSP is not equal to CSP, $P_{j}\ne P_{g}$.
It should be noted that the transmission coefficient $T^{\sigma}$ depends on the Hamiltonian of the device.

For the ferromagnetic device,
the spin wavefunction can be written as
$\chi_{\mu }^{\sigma}=(1+\sigma,1-\sigma)^{\top}/2$ and
$\sigma=\pm1$ labels spin states. The transmission coefficient has a diagonal matrix form,
$T_{f}=diag(T_{f}^{\uparrow },T_{f}^{\downarrow })$, and $T_{f}^{\sigma }=\frac{t_{f}^{\sigma }}{A_{f}^{\sigma }}$.
The BSP in Eq.(\ref{BSP}) can be rewritten as
\begin{equation}
P_{g}^{f}=\frac{Tr(\Omega^{f})}{Tr(D^{f})}
\label{BSPf}
\end{equation}%
where $\Omega^{f}=T_{f}^{\dagger}\sigma_{z}T_{f}$, and $D^{f}=T_{f}^{\dagger}T_{f}$, where
the index $f$ labels the ferromagnetic device.
In order to find out the relationship between the incident and outgoing currents,
we parameterize the amplitudes of the incident wavefunction by $\xi $ and $q$ as
$A_{f}^{\uparrow}=|A_{f}^{\uparrow}|e^{i\xi_{\uparrow}}=e^{i\xi_{\uparrow}}\cos \frac{q}{2}$,
$A_{f}^{\downarrow}=|A_{f}^{\downarrow}|e^{i\xi_{\downarrow}}=e^{i\xi_{\downarrow}}\sin \frac{q}{2}$ and
$A_{f}^{\uparrow *}A_{f}^{\downarrow}=\frac{e^{i\xi}}{2}\sin q$, where $\xi =\xi_{\downarrow}-\xi_{\uparrow}$ and $q\in (0,\pi )$. Thus,
we have $P_{j,in}=\cos q$, namely the CSP of incident current can be described by the parameter $q$. The outgoing CSP can be expressed as (see Appendix A)
\begin{equation}
P_{j,out}^{f}=\frac{Tr[\Omega_{f}(1+\sigma_{n})]}{Tr[D_{f}(1+\sigma_{n}]}
\label{CSPoutf}
\end{equation}%
where $\sigma_{n}=\mathbf{\sigma }\cdot \mathbf{n}$, where
$\mathbf{n}=(\sin q\cos \xi, \sin q\sin \xi, \cos q)$ is a unit vector on $S^{2}$, and
$\mathbf{\sigma}$ is Pauli matrix.

For the SOI semiconductor device, the spin wavefunction of the device can be solved for given Hamiltonian,
but usually $\chi _{c}^{\sigma }\ne(1+\sigma,1-\sigma)^{\top}/2$.
In order to match the wave functions between the device and electrodes, usually
one can make a unitary transformation $U$ from the spinor basis to
the $\sigma_{z}$ basis\cite{Molenkamp,Splettstoesser} to give
the transmission amplitude $t_{s}=Ut\equiv (t_{s}^{\uparrow},t_{s}^{\downarrow})^{\top}$,
and the incident amplitude $A_{s}=UA\equiv (A_{s}^{\uparrow},A_{s}^{\downarrow})^{\top}$,
where the index $s$ labels the SOI semiconductor device.
Thus, the spin-dependent current can be
expressed as $j_{in}^{\sigma }=k|A_{s}^{\sigma }|^{2}$ and $j_{out}^{\sigma }=k|t_{s}^{\sigma }|^{2}$,
where $t_{s}=T_{s}A_{s}$, $T_{s}=UTU^{\dagger }$,
$T=diag(T^{\uparrow },T^{\downarrow })$, and $T^{\sigma }=\frac{t^{\sigma }}{A^{\sigma }}$.
The spin-dependent conductance can be written by $g^{\sigma}=\frac{e^{2}}{h}|T_{s}^{\sigma}|^{2}$, where
$T_{s}^{\uparrow}=T_{s,11}+T_{s,12}$ and $T_{s}^{\downarrow}=T_{s,21}+T_{s,22}$.
Thus, the BSP for SOI semiconductor devices can be obtained as (see Appendix B)
\begin{equation}
P_{g}^{s}=\frac{Tr[\Omega^{s}(1+\sigma_{x})]}{Tr[D^{s}(1+\sigma_{x})]}
\label{BSPs}
\end{equation}%
where $\Omega^{s}=T_{s}^{\dagger}\sigma_{z}T_{s}$, and $D^{s}=T_{s}^{\dagger}T_{s}$.
Similarly,
we parameterize $A^{\sigma}_{s}$ and obtain $P_{j,in}^{s}=\cos q$, where $q\in (0,\pi )$.
Using the spin-dependent outgoing current and the matrix form of the transmission coefficient $T_{s}$,
we can also give the CSP for SOI semiconducting devices, (see Appendix A)
\begin{equation}
P_{j,out}^{s}=\frac{Tr[\Omega^{s}(1+\sigma _{n})]}{Tr[D^{s}(1+\sigma _{n})]},
\label{CSPs}
\end{equation}%
The CSP can be written as a unified form for the ferromagnetic and SOI semiconducting devices, in which
the $\Omega^{f}$ is a diagonal matrix and $\Omega^{s}$ is a non-diagonal matrix.

Generally speaking, BSP is not equal to CSP, $P_{j,out}^{f(s)}\ne P_{g}^{f(s)}$.
We should emphasize that the transmission coefficient $T_{s}$ is an intrinsic variable of devices
and is namely independent of the incident current, while the $t_{s}$ depends on the incident current.
Thus, the BSP measures the intrinsic spin polarization of device, while
the CSP depends on the spin polarization of incident current
and measures the global properties of spin polarization of the whole system.
In practice, the conductance cannot be measured directly, namely BSP cannot be measured directly,
while the CSP can be measured directly because the current can be measured directly.\cite{Kato}

Based on the general formulas in Eqs.(\ref{BSPf})-(\ref{CSPs}),
we can obtain some general theorems on BSP and CSP.

{\bf Theorem I}: For $T^{\sigma}=c\mathcal{T}^{\sigma}$, where $c$ is a complex variable and independent of spin degree freedom,
the $P_{j,out}^{f(s)}$ and $P_{g}^{f(s)}$ are independent of $c$. Namely,
\begin{eqnarray}
&&P_{g}^{f}=\frac{Tr(\bar{\Omega}^{f})}{Tr(\bar{D}^{f})}; \           \
P_{j,out}^{f}=\frac{Tr[\bar{\Omega}_{f}(1+\sigma_{n})]}{Tr[\bar{D}_{f}(1+\sigma_{n}]};\nonumber\\
&&P_{g}^{s}=\frac{Tr[\bar{\Omega}^{s}(1+\sigma_{x})]}{Tr[\bar{D}^{s}(1+\sigma_{x})]}; \  \
P_{j,out}^{s}=\frac{Tr[\bar{\Omega}^{s}(1+\sigma _{n})]}{Tr[\bar{D}^{s}(1+\sigma _{n})]},
\label{CSPs1}
\end{eqnarray}%
where
$\bar{\Omega}_{f(s)}=\mathcal{T}_{f(s)}^{\dagger}\sigma_{z}\mathcal{T}_{f(s)}$,
and $\bar{D}_{f(s)}=\mathcal{T}_{f(s)}^{\dagger}\mathcal{T}_{f(s)}$. (Proof in Appendix C)

{\bf Theorem II}: For $T^{\sigma}=c\mathcal{T}^{\sigma}$, where $c$ is a complex variable,
$diag(\mathcal{T}^{\uparrow},\mathcal{T}^{\downarrow})=\mathcal{T}=\mathcal{T}_{f}$,
$\mathcal{T}_{s}=U\mathcal{T}U^{\dagger}$, $\mathcal{T}_{s}$ and $\mathcal{T}_{f}\in SU(2)$,
then BSP and CSP are obtained (Proof in Appendix D)
\begin{eqnarray}
&&P_{g}^{f}=0; \     \
P_{g}^{s}=\frac{1}{2}Tr[\sigma_{z}\mathcal{T}_{s}\sigma_{x}\mathcal{T}^{\dagger}_{s}],\\
&&P_{j,out}^{f(s)}=\frac{1}{2}Tr[\sigma_{z}\mathcal{T}_{f(s)}\sigma_{n}\mathcal{T}^{\dagger}_{f(s)}],\nonumber
\label{CSPsf2}
\end{eqnarray}

These two theorems can help us to simplify the derivation of BSP and CSP for concretes systems.
We may focus our attention only on the factor of the transmission coefficient
that contributes to BSP and CSP.

{\bf Theorem III}: The relationship between the BSP and CSP can be expressed as,
\begin{equation}
P_{j,out}^{f}=\frac{P_{g}^{f}+P_{j,in}}{1+P_{g}^{f}P_{j,in}}; \     \
P_{j,out}^{s}=\frac{P_{g}^{s}+r_{\Omega}}{1+r_{D}}
\label{BCSPsg}
\end{equation}%
where
$r_{\Omega} =\frac{Tr[\Omega^{s}(\sigma_{n}-\sigma_{x})]}{Tr[D^{s}(1+\sigma_{x})]}$ and
$r_{D} =\frac{Tr[D^{s}(\sigma_{n}-\sigma_{x})]}{Tr[D^{s}(1+\sigma_{x})]}$ (Proof in Appendix E).

We can give the basic relationship between BSP and CSP from the theorem III:
(1) For the non-spin polarized incident current, $P_{j,in}=0$, $P_{j}^{f}=P_{g}^{f}$ and $\sigma_{n}=\sigma_{x}\cos\xi+\sigma_{y}\sin\xi$,
if $\xi=2n\pi$, where $n$ is an integer, we have $P_{j,out}^{s}=P_{g}^{s}$;
(2) for the full-spin-polarized incident current,$P_{j,in}=1$,$\sigma_{n}=\sigma_{z}$, we have $P_{j}^{f}=1$, and
$r_{\Omega} =\frac{Tr[\Omega^{s}(\sigma_{z}-\sigma_{x})]}{Tr[D^{s}(1+\sigma_{x})]}$ and
$r_{D} =\frac{Tr[D^{s}(\sigma_{z}-\sigma_{x})]}{Tr[D^{s}(1+\sigma_{x})]}$.

In general, the spin wavefunction for the SOI semiconducting device can be written in the spinor representation,
$\chi^{\uparrow}=(\cos\theta/2,-e^{i\varphi}\sin\theta/2)^{\top}$
and $\chi^{\downarrow}=(\sin\theta/2,e^{i\varphi}\cos\theta/2)^{\top}$.
The unitary transformation can be given as
\begin{equation}
U=\left(
\begin{array}{cc}
\cos\theta/2 & \sin\theta/2 \\
-e^{i\varphi}\sin\theta/2& e^{i\varphi}\cos\theta/2,
\end{array}%
\right)
\end{equation}
where $\theta$ and $\varphi$ are parameters depending on the Hamiltonian of the device.
Suppose that $T^{\sigma}=\tau^{\sigma}e^{i\phi^{\sigma}}$ we can give the BSP from Eqs. (\ref{BSPf}) and (\ref{CSPoutf})
\begin{equation}
P_{g}^{f}=\frac{\tau_{\uparrow}^{2}-\tau_{\downarrow}^{2}}
{\tau_{\uparrow}^{2}+\tau_{\downarrow}^{2}}
\label{Pgf}
\end{equation}
\begin{equation}
P_{j,out}^{f}=\frac{\tau_{\uparrow}^{2}-\tau_{\downarrow}^{2}+(\tau_{\uparrow}^{2}+\tau_{\downarrow}^{2})\cos q}
{\tau_{\uparrow}^{2}+\tau_{\downarrow}^{2}+(\tau_{\uparrow}^{2}-\tau_{\downarrow}^{2})\cos q}
\label{Pjf}
\end{equation}
Using Eqs.(\ref{BSPs}) and (\ref{CSPs}), we can obtain the CSP, (Derivation in Appendix F)
\begin{equation}
P_{g}^{s}=\frac{\tau_{-}^{2} \cos \theta
-\tau_{+}^{2}b_{g}\cos \theta +2\tau^{\uparrow }\tau^{\downarrow}a_{g}\sin \theta}
{\tau_{+}^{2}-\tau_{-}^{2}\sin \theta \cos \varphi},
\label{Pgs}
\end{equation}
\begin{equation}
P_{j,out}^{s}=\frac{\tau_{-}^{2}\cos \theta
-\tau_{+}^{2}b_{j}\cos\theta+2\tau^{\uparrow }\tau^{\downarrow}a_{j}\sin \theta }
{\tau_{+}^{2}+\tau_{-}^{2}[\cos\theta\cos q-\sin\theta\sin q\cos(\varphi-\xi)]}
\label{Pjs}
\end{equation}
where
\begin{equation}
\left(
\begin{array}{c}
a_{g}\\
b_{g}
\end{array}
\right)=
\left(
\begin{array}{c}
\cos\Delta\phi\cos \theta \cos \varphi-\sin\Delta\phi\sin \varphi\\
\sin \theta\cos \varphi
\end{array}
\right)
\label{abg}
\end{equation}
\begin{equation}
\left(
\begin{array}{c}
a_{j}\\
b_{j}
\end{array}
\right)=
\left(
\begin{array}{c}
u\cos\Delta\phi-\sin\Delta\phi \sin q\sin(\varphi-\xi)\\
\sin\theta \sin q \cos(\varphi-\xi)-\cos\theta \cos q
\end{array}
\right)
\label{abj}
\end{equation}
and $\tau_{\pm}^{2}=\tau^{\uparrow 2} \pm\tau^{\downarrow 2}$,
$\Delta\phi=\phi^{\uparrow}-\phi^{\downarrow}$, and $u=\cos \theta \sin q \cos (\varphi-\xi)+\sin \theta \cos q$.
The general formulas of BSP and CSP in Eqs. (\ref{Pgf})-(\ref{Pjs}) describe the basic properties of
the spin-dependent electron transport in mesoscopic ferromagnetic and SOI semiconductor systems.
The BSP describes the intrinsic spin polarized properties of devices, while the CSP
describes the global spin polarized properties of the whole system, including the incident-current
amplitude and phase differences.
In this spinor representation, we can give
\begin{equation}
\left(
\begin{array}{c}
r_{D}\\
r_{\Omega}
\end{array}
\right)=
\frac{1}{\omega}\left(
\begin{array}{c}
\tau_{-}^{2}(\sin\theta \cos\varphi-b_{j})\\
\tau_{+}^{2}c_{1}\cos\theta +2\tau^{\uparrow }\tau^{\downarrow}c_{2}\sin \theta
\end{array}
\right)
\label{rr}
\end{equation}
where
$\omega=\tau_{+}^{2}-\tau_{-}^{2}\sin \theta \cos\varphi$
and
\begin{equation}
\left(
\begin{array}{c}
c_{1}\\
c_{2}
\end{array}
\right)=
\left(
\begin{array}{c}
v_{1}\sin\theta+\cos q\cos\theta\\
\cos\Delta\phi[\cos q\sin\theta-v_{1}\cos\theta]
+v_{2}\sin\Delta\phi
\end{array}
\right)
\label{cc}
\end{equation}
where $v_{1}=\cos\varphi-\sin q\cos(\varphi-\xi)$, and $v_{2}=\sin\varphi-\sin(\varphi-\xi)\sin q$.
To further study the basic properties
of spin polarization in mesoscopic SOI systems, we analyze several typical cases.

{\bf Case 1}: For the non-spin polarized incident currents, $P_{j,in}=0$,
the CSP in Eq. (\ref{Pjs}) becomes
\begin{equation}
P_{j,out}^{s}=\frac{\tau_{-}^{2}\cos \theta
-\tau_{+}^{2}b_{j}\cos\theta+2\tau^{\uparrow }\tau^{\downarrow}a_{j}\sin \theta }
{\tau_{+}^{2}-\tau_{-}^{2}\sin\theta\cos(\varphi-\xi)}
\label{Pjs1}
\end{equation}
where
\begin{equation}
\left(
\begin{array}{c}
a_{j}\\
b_{j}
\end{array}
\right)=
\left(
\begin{array}{c}
\cos\Delta\phi\cos \theta \cos (\varphi-\xi)-\sin\Delta\phi\sin(\varphi-\xi)\\
\sin\theta \cos(\varphi-\xi)
\end{array}
\right)
\label{abj0}
\end{equation}
\begin{equation}
\left(
\begin{array}{c}
c_{1}\\
c_{2}
\end{array}
\right)=
\left(
\begin{array}{c}
v_{3}\sin\theta\\
v_{4}\sin\Delta\phi-v_{3}\cos\Delta\phi\cos\theta.
\end{array}
\right)
\label{cc0}
\end{equation}
where $v_{3}=\cos\varphi-\cos(\varphi-\xi)$, and $v_{4}=\sin\varphi-\sin(\varphi-\xi)$.
It can be seen that $P_{g}^{s}\ne P_{j,out}^{s}$ generally even when $P_{j,in}=0$
because of the phase effect of incident current $\xi$. This is different
from the ferromagnetic device. For the in-phase cases, $\xi=2n\pi$ where $n$ is an integer,
$c_{1}=c_{2}=0$, and $r_{\Omega}=r_{D}=0$. Hence $P_{j,out}^{s}=P_{g}^{s}$.

{\bf Case 2}: For the full-spin polarized incident current, $P_{j,in}=1$, the CSP can be reduced to
\begin{equation}
P_{j,out}^{s}=\frac{\tau_{-}^{2}\cos \theta
+\tau_{+}^{2}\cos^{2}\theta
+2\tau^{\uparrow }\tau^{\downarrow}\cos\Delta\phi^{\sigma}\sin^{2}\theta}
{\tau_{+}^{2}+\tau_{-}^{2}\cos\theta},
\label{Pjs2}
\end{equation}
\begin{equation}
\left(
\begin{array}{c}
c_{1}\\
c_{2}
\end{array}
\right)=
\left(
\begin{array}{c}
\sin\theta\cos\varphi+\cos\theta\\
\cos\Delta\phi(\sin\theta-\cos\varphi\cos\theta)
+\sin\Delta\phi\sin\varphi
\end{array}
\right)
\label{cc1}
\end{equation}

{\bf Case 3}: $T^{\sigma}=\tau e^{i\phi^{\sigma}}$, namely $\tau_{-}=0$, the BSP and CSP in Eqs. (\ref{Pgs}) and (\ref{Pjs})
can be simplified to,
\begin{equation}
P_{g}^{s}=a_{g}\sin \theta-b_{g}\cos \theta
\label{Pgs3}
\end{equation}
\begin{equation}
P_{j,out}^{s}=a_{j}\sin \theta-b_{j}\cos \theta
\label{Pjs3}
\end{equation}
where $a_{g(j)}$ and $b_{g(j)}$ are same to Eqs.(\ref{abg}) and (\ref{abj})
since they are independent of $\tau^{\sigma}$. The $r_{D}=0$ and
$r_{\Omega}=c_{1}\cos\theta +c_{2}\sin \theta$, where $c_{1(2)}$ is same to Eq. (\ref{cc}). Actually,
many concrete systems belong to this case.
For the non-spin polarized incident current, $P_{j,in}=0$,
the $a_{j}$, $b_{j}$ and $c_{1(2)}$ reduce to Eqs.(\ref{abj0}) and (\ref{cc0}).
For the full-spin polarized incident currents, $P_{j,in}=1$,
$P_{j,out}^{s}=\cos\Delta\phi\sin^{2}\theta+\cos^{2}\theta$.
The $c_{1(2)}$ reduce to Eq. (\ref{cc1}).
Similarly, $P_{g}^{s}\ne P_{j}^{s}$ generally. The spin-phase coherence
of the incident current $\xi$ plays some roles in CSP.

{\bf Case 4}: $T^{\sigma}=\tau^{\sigma}e^{i\phi}$, namely $\Delta\phi=0$,
the $P_{g}^{s}$ and $P_{j}^{s}$ in Eqs.(\ref{Pgs}) and (\ref{Pjs}) still hold, where
\begin{equation}
\left(
\begin{array}{c}
a_{g}\\
b_{g}
\end{array}
\right)=
\left(
\begin{array}{c}
\cos \theta \cos\varphi\\
\sin\theta \cos\varphi
\end{array}
\right);
\label{abg2}
\end{equation}
\begin{equation}
\left(
\begin{array}{c}
a_{j}\\
b_{j}
\end{array}
\right)=
\left(
\begin{array}{c}
\cos \theta \sin q \cos (\varphi-\xi)+\sin \theta \cos q\\
\sin\theta \sin q \cos(\varphi-\xi)-\cos\theta \cos q
\end{array}
\right);
\label{abj2}
\end{equation}
\begin{equation}
\left(
\begin{array}{c}
c_{1}\\
c_{2}
\end{array}
\right)=
\left(
\begin{array}{c}
\sin\theta[\cos\varphi-\sin q\cos(\varphi-\xi)]+\cos q\cos\theta\\
\cos q\sin\theta+[\cos(\varphi-\xi)\sin q-\cos\varphi]\cos\theta
\end{array}
\right).
\label{cc2}
\end{equation}

For the non-spin polarized incident currents, $P_{j,in}=0$,
the $P_{j,out}^{s}$ also reduces the Eq. (\ref{Pjs1}), where
$a_{j}=\cos \theta \cos(\varphi-\xi)$;
$b_{j}=\sin \theta \cos(\varphi-\xi)$;
$c_{1}=\sin\theta[\cos\varphi-\cos(\varphi-\xi)]$, and
$c_{2}=[\cos(\varphi-\xi)-\cos\varphi]\cos\theta$.
For the full-spin polarized incident currents, $P_{j,in}=1$,
the $P_{j,out}^{s}$ is same to the Eq. (\ref{Pjs2}), where
$c_{1}=\sin\theta\cos\varphi+\cos\theta$ and
$c_{2}=\sin\theta-\cos\varphi\cos\theta$.
This formalism gives the basic properties of BSP and CSP, and their relationship.

\section{Typical device models}
\subsection{Ferromagnetic nano wires}
We first consider a ferromagnetic one-dimensional nano wire. The reduced
Hamiltonian can be written as
\begin{equation}
\mathcal{H}=k_{x}^{2}+\sigma\Delta_{F}
\label{Fwire}
\end{equation}%
where $\sigma=\pm 1$ for spin-up and -down states and $\Delta_{F}$ is the spin-split
energy gap. The eigenenergy can be given as $\varepsilon _{\sigma
}=k_{x}^{2}+\sigma \Delta_{F}$, and its corresponding spin wavefunction is
$\chi _{c}^{\sigma }=(1+\sigma,1-\sigma)^{\top}/2$.
Using the wavefunction in Eq. (\ref%
{wave}) with the spin-dependent Griffth boundary condition\cite{Griffith}, we obtain the
transmission coefficient,
\begin{equation}
T^{\sigma}=\frac{e^{-ik_{F}L}}{\cos (k^{\sigma}L)-\overline{\kappa}^{\sigma}_{F}\sin(k^{\sigma}L)}
\label{Tfwire}
\end{equation}
where $L$ is the length of the central nanowire device, and
$\overline{\kappa}^{\sigma}_{F}=\frac{1}{2}(\frac{k^{\sigma}}{k_{F}}+\frac{k_{F}}{k^{\sigma}})$,
and $k^{\sigma}=\sqrt{k_{F}^{2}-\sigma\Delta_{F}}$.
Thus, we can obtain $P_{g}^{f}$ from Eq. (\ref{BSPf}).
\begin{equation}
P_{g}^{f}=\frac{C_{s,-}+S_{n,-}+S_{n2}}{C_{s,+}+S_{n,+}-S_{n2}}
\end{equation}
where
\begin{eqnarray}
C_{s,\pm}&=&\cos^{2}(k^{\downarrow}L)\pm\cos^{2}(k^{\uparrow}L)\nonumber\\
S_{n,\pm}&=&\overline{\kappa}^{\downarrow2}_{F}\sin^{2}(k^{\downarrow}L)\pm
\overline{\kappa}^{\uparrow2}_{F}\sin^{2}(k^{\uparrow}L) \\
S_{n2}&=&\overline{\kappa}^{\uparrow}_{F}\sin(2k^{\uparrow}L)-
\overline{\kappa}^{\downarrow}_{F}\sin(2k^{\downarrow}L).\nonumber
\end{eqnarray}
The CSP $P_{j,out}^{f}$ can be obtained from Eq. (\ref{BCSPsg}) straightforwardly.

\subsection{Semiconducting nano wires}
The second typical example is the one-dimensional Rashba
SOI nanowire. The reduced Hamiltonian can be written as \cite{Molenkamp}%
\begin{equation}
\mathcal{H}=(p_{x}-\frac{\overline{\alpha }}{2}\sigma _{y})^{2}  \label{wireH}
\end{equation}%
where
$\overline{\alpha }=\frac{2\alpha m}{\hbar ^{2}}$ is the reduced Rashba
strength. The eigenenergy can be given as $\varepsilon _{\sigma
}=(k_{x}+\sigma \frac{\overline{\alpha }}{2})^{2}$, and its corresponding wavefunction is
$\psi _{n\lambda }^{\sigma }=\frac{1}{\sqrt{2\pi }}%
e^{ik_{\lambda }^{\sigma }x}\chi _{c}^{\sigma }$, where the spinor wavefunction is $\chi
_{c}^{\sigma }=\frac{1}{\sqrt{2}}\left(
1 , -\sigma i\right)^T $, where $\sigma =\pm $ labels spin states and $\lambda =\pm $
labels the direction of electron motion. Similarly, we can obtain the
transmission coefficient in the spinor basis,
$T^{\sigma }=ce^{i\sigma \overline{\alpha }L/2}$
where
$c=\frac{e^{-ik_{F}L}}{\cos(kL)-i\overline{\kappa}\sin(kL)}$
where $\overline{\kappa}=\frac{1}{2}(\frac{k}{k_{F}}+\frac{k_{F}}{k})$ and
$k=\sqrt{k_{F}^{2}+\frac{\overline{\alpha}^{2}}{4}}$.
Notice that $\theta=\varphi=\pi/2$ and $\Delta\phi=\overline{\alpha }L$,
the CSP can be obtained
\begin{equation}
P_{j,out}^{s}=\cos \overline{\alpha }L\cos q+P_{g}^{s}\cos \xi \sin q,
\label{Pjswire}
\end{equation}%
where the BSP is
$P_{g}^{s}=-\sin(\overline{\alpha }L)$.
It can be seen that BSP $P_{g}^{s}$ depends only on the intrinsic variables of the device,
reduced Rashba constant $\overline{\alpha}$ and the length of nanowire $L$, while
the CSP $P_{j}^{s}$ depends on both of the intrinsic variables and the incident current.

For the non-spin polarized incident current $\cos q=0$,
$P_{j,out}^{s}=P_{g}^{s}\cos \xi$. Namely, the spin-phase coherence $\xi$ modulates the CSP.
If $\xi=2n\pi$, $P_{j}^{s}=P_{g}^{s}$.
For the full-spin-polarized incident current $P_{j,in}=1$, $P_{j,out}^{s}=\cos(\overline{\alpha }L)$, which
has $-\pi/2$ phase shift to $P_{g}^{s}$.
For the cases of in-phase or anti-phase, namely $\xi=2n\pi$ or $\xi=(2n+1)\pi$, $P_{j,out}^{s}=cos(\overline{\alpha }L\pm q)$.
Interestingly, these results imply that the non-spin-polarized incident current can be polarized by the
Rashba SOI nanowire. Reversely, the spin-polarized incident current can be destroyed by some specific
parameters $\xi,\overline{\alpha }$, and $L$. These results provide us a method to control the CSP $P_{j}^{s}$ by
designing the length $L$ and tuning the SOI strength of the nanowire.

\subsection{Semiconducting nano rings}
The third typical nano device model is the mesoscopic Rashba ring with two electrodes. Its reduced Hamiltonian is\cite{Meijer,Citro}%
\begin{equation}
\mathcal{H}=(-i\frac{\partial }{\partial \varphi }+\frac{\overline{\alpha }R}{2}\sigma _{r})^{2}
\end{equation}%
where $\sigma _{r}=\sigma _{x}\cos \varphi +\sigma _{y}\sin \varphi$, $\overline{\alpha }=\frac{2\alpha m^*}{\hbar^{2}}$
is the reduced Rashba strength, and $R$ is the radius of the ring.
The eigenenergy $\varepsilon _{n\lambda }^{\sigma }=(n_{\lambda }^{\sigma }+%
\frac{\Phi _{AC}^{\sigma }}{2\pi })^{2}$, where $\Phi _{AC}^{\sigma }=-\pi
(1-\sigma \sqrt{1+\bar{\alpha}^{2}R^{2}})$ is the AC phase and  $n_{\lambda
}^{\sigma }$ is determined by the energy matching between the ring and leads, which satisfy
$n_{+}^{\sigma }+n_{-}^{\sigma }=\frac{\Phi _{AC}^{\sigma }}{\pi }$ and
$n_{+}^{\sigma }-n_{-}^{\sigma }=2kR$.
The wavefunction is $\psi _{n\lambda }^{\sigma }=\frac{1}{\sqrt{2\pi }}%
e^{in_{\lambda }^{\sigma }\varphi }\chi ^{\sigma }(\varphi )$, where $\chi
^{\sigma }(\varphi )$ is the spinor wavefunction.\cite{Huang}
$\chi ^{\sigma }=\left(
\sin \frac{2\theta +\pi (\sigma +1)}{4},
e^{i\varphi }\cos \frac{2\theta +\pi (\sigma +1)}{4}
\right)^{\top}$,
where $\lambda =\pm 1$ represents electronic moving direction. We consider
that the angle between two electrodes is $\pi$. Using the similar method, the
transmission coefficient can be expressed in terms of
$T^{\sigma}=c\mathcal{T}^{\sigma }$,\cite{Huang,Aronov,Molnar}  where
\begin{equation}
c=\frac{4i\sin(kR\pi)\sin(\Delta_{AC}/2)}{ \sin^{2}(kR\pi)-2\cos\Delta _{AC}-2e^{-i2kR\pi }}
\end{equation}
is the spin-independent factor,
and $\Delta_{AC}=\pi \sqrt{1+\bar{\alpha}^{2}R^{2}}$.
The spin-dependent factor is $\mathcal{T}^{\sigma }=\sigma$.
It should be noted that $T_{s}=U(\varphi)TU(0)^{\dagger}$ for the ring device, where $\varphi$ is the
angle between two leads. For the specific case, $\varphi=\pi$, we have
$\Omega^{s}=\sigma_{z}\cos2\theta-\sigma_{x}\sin2\theta$.
The CSP can be obtained
\begin{equation}
P_{j,out}^{s}=\cos 2\theta \cos q+P_{g}^{s}\sin q\cos \xi
\label{Pjsring}
\end{equation}%
where $P_{g}^{s}=-\sin2\theta $. For the non-spin polarized incident current $P_{j,in}=0$,
$P_{j,out}^{s}=P_{g}^{s}\cos\xi$. Similarly to the nanowire case, the spin-phase coherence $\xi$ of incident current
modulates the CSP. When $\xi=2n\pi$, $P_{j,out}^{s}=P_{g}^{s}$.
For full-spin polarized incident current, $P_{j,in}=1$, the CSP,
$P_{j,out}^{s}=\cos(2\theta)$, has a $-\pi/2$ phase shift to BSP.
For the in-phase or anti-phase case, $\xi =2n\pi$ or $(2n+1)\pi$,
$P_{j,out}^{s}=\cos (2\theta \pm q)$. For $\xi =(2n+1)\pi /2$,
$P_{j,out}^{s}=\cos 2\theta \cos q$.

Notice that $\tan \theta =\bar{\alpha}R$, $P_{j,out}^{s}$ depends on the radius of ring $R$, the SOI strength $\bar{\alpha}$,
and the incident current parameters $\xi$\ and $q$. For a given incident current, $\xi $ and $q$, the $P_{j,out}^{s}$
can be tuned by the SOI strength $\bar{\alpha}$ and the geometric parameters of the ring $R$. Thus, we have quite large space
to tune the spin polarization of outgoing current by designing the geometry of the ring and tuning the SOI
strength through applied electric field. Similarly, the incident current
can be polarized and also destroyed for some specific parameters.

\section{Discussion and conclusion}
It should be pointed out that the basic results are qualitatively invariant for
Dresselhaus SOI and the finite width belts and rings. For the Dresselhaus
ring, it has been found that the transport properties of the spin-dependent
currents are very similar to that of the Rashba ring.\cite{Molenkamp}
For the finite-width belts and rings,\cite{Frustaglia} mobile electrons
have many transport channels that modify the detail properties we obtain here,
but the CSP of incident current can be still tuned by the finite-width belts and rings even though we
cannot obtain the analytic formula of $P_{j,out}^{s}$.
Interestingly, for many-electron systems, when the incident current is spin-phase-coherent, namely the
spin-phase difference $\langle{\cos\xi}\rangle\neq 0$, the interference effect induced by the spin-phase
coherence modulates the CSP. For the non-spin-coherent incident current, $\langle{\cos\xi}\rangle=0$,
the spin-phase effect of the incident current vanishes.

In summary, we establish a general formalism of BSP and CSP, which is applicable for both mesoscopic
ferromagnetic and SOI semiconducting systems. Based on the formalism, we reveal the basic properties
of BSP and CSP and their relationships, which clearly indicates that they are in general different and reflecting
the different physical aspects. The BSP depends only on the intrinsic variables of the device,
such as the geometric parameter of device and SOI strength or ferromagnetic parameter of device.
Thus, the BSP describes the intrinsic spin polarized properties of the device. The CSP depends on both
of intrinsic parameters of device and the incident current. It describes the global spin-polarized properties
of whole system. The spin-phase coherence of the incident current plays an interference effect and modulates the CSP.
For the non-spin polarized incident current with the in-phase spin-phase coherence, CSP equals to BSP.
The analytical BSP and CSP of several typical nano device models,
ferromagnetic nano wire, Rashba nano wire and rings, exhibit basic physical behaviors of BSP and CSP and their relationships.
Our study of spin polarization helps us to identify a fact that, when we design a spintronics device,
we have to consider the global effect which is obviously important in integrated circuit and these results also
provide some hints to design practical spintronics devices.

{\bf Acknowledgments:}
We thank Paul Erdos for helpful discussions. This work was supported financially by the
Fundamental Research Funds for the Central Universities.


\section{Appendixes}


\subsection{Derivation of Eqs. (\ref{CSPoutf}) and (\ref{CSPs})}
Notice that for any $2\times 2$ matrix $X$ and $2$-row vector $\alpha$, we have
$\alpha^{\dagger}X\alpha=Tr(X\alpha\otimes\alpha^{\dagger})$.
For $\alpha_{f(s)}=(A^{\uparrow}_{f(s)},A^{\downarrow}_{f(s)})^{\top}$,where
$A^{\uparrow}_{f(s)}=|A^{\uparrow}_{f(s)}|e^{i\xi_{\uparrow}}=e^{i\xi_{\uparrow}}\cos \frac{q}{2}$,
and $A^{\downarrow}_{f(s)}=|A^{\downarrow}_{f(s)}|e^{i\xi_{\downarrow}}=e^{i\xi_{\downarrow}}\sin\frac{q}{2}$,
we can obtain $\alpha_{f(s)}\otimes\alpha^{\dagger}_{f(s)}=(1+\sigma_{n})/2$. Thus,
\begin{eqnarray}
P_{j,out}^{f(s)}&=&\frac{|T^{\uparrow}_{f(s)}|^{2}|A^{\uparrow}_{f(s)}|^{2}-|T^{\downarrow}_{f(s)}|^{2}|A^{\downarrow}_{f(s)}|^{2}}
{|T^{\uparrow}_{f(s)}|^{2}|A^{\uparrow}_{f(s)}|^{2}+|T^{\downarrow}_{f(s)}|^{2}|A^{\downarrow}_{f(s)}|^{2}}\nonumber\\
&=&\frac{\alpha^{\dagger}_{f(s)}\Omega^{f(s)}\alpha_{f(s)}}{\alpha^{\dagger}_{f(s)}D^{f(s)}\alpha_{f(s)}}\nonumber\\
&=&\frac{Tr(\Omega^{f(s)}\alpha_{f(s)}\otimes\alpha^{\dagger}_{f(s)})}{Tr(D^{f(s)}\alpha_{f(s)}\otimes\alpha^{\dagger}_{f(s)})}\nonumber\\
&=&\frac{Tr[\Omega^{f(s)}(1+\sigma_{n})]}{Tr[D^{f(s)}(1+\sigma_{n})]}.
\label{CSPpf}
\end{eqnarray}%
where $\Omega^{f(s)}=T_{f(s)}^{\dagger}\sigma_{z}T_{f(s)}$, and $D^{f(s)}=T_{f(s)}^{\dagger}T_{f(s)}$.

\subsection{Derivation of Eq. (\ref{BSPs})}
Notice that $T^{\uparrow}_{s}=T_{s,11}+T_{s,12}$, $T^{\downarrow}_{s}=T_{s,21}+T_{s,22}$,
$\Omega^{s}=T^{\dagger}_{s}\sigma_{z}T_{s}$ and $D^{s}=T^{\dagger}_{s}T_{s}$, we have
\begin{eqnarray}
P_{g}^{s} &=& \frac{|T^{\uparrow}_{s}|^{2}-|T^{\downarrow}_{s}|^{2}}{|T^{\uparrow}_{s}|^{2}+|T^{\downarrow}_{s}|^{2}}
=\frac{\sum_{i,j}\Omega^{s}_{ij}}{\sum_{i,j}D^{s}_{ij}}\nonumber\\
&=& \frac{Tr[\Omega^{s}(1+\sigma_{x})]}{Tr[D^{s}(1+\sigma_{x})]},
\end{eqnarray}

\subsection{Proof of theorem I}
Proof: For $T_{f(s)}^{\sigma}=c\mathcal{T}_{f(s)}^{\sigma}$, where $c$ is a complex variable,
$\Omega^{f(s)}=T_{f(s)}^{\dagger}\sigma_{z}T_{f(s)}=
|c|^{2}\mathcal{T}_{f(s)}^{\dagger}\sigma_{z}\mathcal{T}_{f(s)}$,
and $D^{f(s)}=T_{f(s)}^{\dagger}T_{f(s)}=|c|^{2}\mathcal{T}_{f(s)}^{\dagger}\mathcal{T}_{f(s)}$.
Substituting the two expressions into Eqs.(\ref{BSPs}) and (\ref{CSPs}) and canceling
$|c|^{2}$, we can obtain Eq. (\ref{CSPs1})

\subsection{Proof of theorem II}
Proof: For $T^{\sigma}=c\mathcal{T}^{\sigma}$, where $c$ is a complex variable;
$\mathcal{T}_{s}$ and $\mathcal{T}_{f}\in SU(2)$,
Notice that $\Omega^{f(s)}=T_{f(s)}^{\dagger}\sigma_{z}T_{f(s)}=
|c|^{2}\mathcal{T}_{f(s)}^{\dagger}\sigma_{z}\mathcal{T}_{f(s)}$,
and $D^{f(s)}=T_{f(s)}^{\dagger}T_{f(s)}=|c|^{2}\mathcal{T}_{f(s)}^{\dagger}\mathcal{T}_{f(s)}=|c|^{2}I$.
\begin{eqnarray}
P^{f}_{g}&=&\frac{Tr(\Omega^{f})}{Tr(D_{f})}
=\frac{|c|^{2}Tr(\mathcal{T}_{f}^{\dagger}\sigma_{z}\mathcal{T}_{f})}
{|c|^{2}Tr(\mathcal{T}_{f}^{\dagger}\mathcal{T}_{f})}\nonumber\\
&=& \frac{Tr(\sigma_{z}\mathcal{T}_{f}\mathcal{T}_{f}^{\dagger})}{2}=\frac{Tr(\sigma_{z})}{2}=0.
\end{eqnarray}
\begin{eqnarray}
P_{g}^{s} &=&\frac{Tr[\Omega^{s}(1+\sigma_{x})]}{Tr[D^{s}(1+\sigma_{x})]}\nonumber\nonumber\\
&=& \frac{|c|^{2}Tr[\mathcal{T}_{s}^{\dagger}\sigma_{z}\mathcal{T}_{s}(1+\sigma_{x})]}
{|c|^{2}Tr[\mathcal{T}_{s}^{\dagger}\mathcal{T}_{s}(1+\sigma_{x})]}\nonumber\\
&=& \frac{Tr[\sigma_{z}\mathcal{T}_{s}(1+\sigma_{x})\mathcal{T}_{s}^{\dagger}]}
{Tr[\mathcal{T}_{s}^{\dagger}\mathcal{T}_{s}]+Tr[\mathcal{T}_{s}^{\dagger}\mathcal{T}_{s}\sigma_{x}]}\nonumber\\
&=& \frac{Tr[\sigma_{z}\mathcal{T}_{s}\mathcal{T}_{s}^{\dagger}]
+Tr[\sigma_{z}\mathcal{T}_{s}\sigma_{x}\mathcal{T}_{s}^{\dagger}]}
{Tr[\mathcal{T}_{s}^{\dagger}\mathcal{T}_{s}]}\nonumber\\
&=& \frac{Tr[\sigma_{z}\mathcal{T}_{f(s)}\sigma _{x}\mathcal{T}^{\dagger}_{s}]}{2},
\end{eqnarray}
\begin{eqnarray}
P_{j,out}^{s} &=&\frac{Tr[\Omega^{s}(1+\sigma_{n})]}{Tr[D^{s}(1+\sigma_{n})]}\nonumber\nonumber\\
&=& \frac{|c|^{2}Tr[\mathcal{T}_{s}^{\dagger}\sigma_{z}\mathcal{T}_{s}(1+\sigma_{n})]}
{|c|^{2}Tr[\mathcal{T}_{s}^{\dagger}\mathcal{T}_{s}(1+\sigma_{n})]}\nonumber\\
&=& \frac{Tr[\sigma_{z}\mathcal{T}_{s}(1+\sigma_{n})\mathcal{T}_{s}^{\dagger}]}
{Tr[\mathcal{T}_{s}^{\dagger}\mathcal{T}_{s}]+Tr[\mathcal{T}_{s}^{\dagger}\mathcal{T}_{s}\sigma_{n}]}\nonumber\\
&=& \frac{Tr[\sigma_{z}\mathcal{T}_{s}\mathcal{T}_{s}^{\dagger}]
+Tr[\sigma_{z}\mathcal{T}_{s}\sigma_{n}\mathcal{T}_{s}^{\dagger}]}
{Tr[\mathcal{T}_{s}^{\dagger}\mathcal{T}_{s}]}\nonumber\\
&=& \frac{Tr[\sigma_{z}\mathcal{T}_{f(s)}\sigma _{n}\mathcal{T}^{\dagger}_{s}]}{2},
\end{eqnarray}

\subsection{Proof of theorem III}
Proof:
For ferromagnetic devices, the CSP can be written as
\begin{eqnarray}
P_{j,out}^{f}&=&\frac{Tr[\Omega^{f}(1+\sigma _{n})]}{Tr[D^{f}(1+\sigma _{n})]}\nonumber\\
&=&\frac{Tr(\Omega^{f})+Tr(\Omega^{f}\sigma _{n})}{Tr(D^{f})+Tr(D^{f}\sigma _{n})}\nonumber\\
&=&\frac{Tr(\Omega^{f})/Tr(D^{f})+Tr(\Omega^{f}\sigma _{n})/Tr(D^{f})}{1+Tr(D^{f}\sigma _{n})/Tr(D^{f})},
\end{eqnarray}
Notice that $Tr(\Omega^{f}\sigma _{n})=(T^{\uparrow 2}+T^{\downarrow 2})\cos q$, $Tr(D^{f})=T^{\uparrow 2}+T^{\downarrow 2}$,
and $Tr(D^{f}\sigma _{n})=Tr(\Omega^{f})\cos q$, we obtain
\begin{eqnarray}
P_{j,out}^{f}&=&\frac{P_{g}^{f}+\cos q}{1+P_{g}^{f}\cos q}\nonumber\\
&=&\frac{P_{g}^{f}+P_{j,in}}{1+P_{g}^{f}P_{j,in}}
\end{eqnarray}

For SOI semiconducting devices,  the $\Omega^{s}$ and $D^{s}$ are not diagonal matrix.
We can separate the trace of the matrix
\begin{eqnarray}
P_{j,out}^{s}&=&\frac{Tr[\Omega^{s}(1+\sigma _{n})]}{Tr[D^{s}(1+\sigma _{n})]}\nonumber\\
&=&\frac{Tr[\Omega^{s}(1+\sigma _{x})]+Tr[\Omega^{s}(\sigma _{n}-\sigma_{x})]}
{Tr[D^{s}(1+\sigma _{x})]+Tr[D^{s}(\sigma _{n}-\sigma_{x})]}\nonumber\\
&=& \frac{P_{g}^{s}+r_{\Omega}}{1+r_{D}},
\end{eqnarray}
where $P_{g}^{s}=\frac{Tr[\Omega^{s}(1+\sigma_{x})]}{Tr[D^{s}(1+\sigma _{x})]}$,
$r_{\Omega}=\frac{Tr[\Omega^{s}(\sigma _{n}-\sigma_{x})]}{Tr[D^{s}(1+\sigma _{x})]}$, and
$r_{D}=\frac{Tr[D^{s}(\sigma _{n}-\sigma_{x})]}{Tr[D^{s}(1+\sigma _{x})]}$.

\subsection{Derivation of Eqs. (\ref{Pgs}) and (\ref{Pjs})}
In general, for Rashba and Dresslhous models, the spin wavefunction can expressed in terms of
the spinor basis,
$\chi^{\uparrow}=(\cos\theta/2,-e^{i\varphi}\sin\theta/2)^{\top}$
and $\chi^{\downarrow}=(\sin\theta/2,e^{i\varphi}\cos\theta/2)^{\top}$.
The transmission coefficient is
$T_{s}=UTU^{\dagger}=T^{\uparrow}\mu_{1}+T^{\downarrow}\mu_{2}
+R\mu_{3}$,
where
$R=\frac{1}{2}(T^{\uparrow}-T^{\downarrow})\sin \theta$ and
\begin{eqnarray}
\mu_{1}&=&\left(
\begin{array}{cc}
\cos^{2}\theta/2 & 0 \\
0 & \sin^{2}\theta/2
\end{array}
\right);\nonumber\\
\mu_{2}&=&\left(
\begin{array}{cc}
\sin^{2}\theta/2 & 0 \\
0 & \cos^{2}\theta/2
\end{array}
\right);\\
\mu_{3}&=&\left(
\begin{array}{cc}
0 & e^{-i\varphi} \\
e^{i\varphi} & 0
\end{array}
\right);\nonumber
\end{eqnarray}
\begin{equation}
U=\left(
\begin{array}{cc}
\cos\theta/2 & \sin\theta/2 \\
-e^{i\varphi}\sin\theta/2& e^{i\varphi}\cos\theta/2
\end{array}%
\right).
\end{equation}
It is straightforward that we have
\begin{equation}
\Omega_{s}=T^{\dagger}_{s}\sigma_{z}T_{s}
=|T^{\uparrow}|^{2}\varsigma_{1-}+|T^{\downarrow}|^{2}\varsigma_{2-}-Z_{-}\sigma_{z}
+d_{1}\varpi_{1-}+d_{2}\varpi_{2-},
\end{equation}
\begin{equation}
D_{s}=T^{\dagger}_{s}T_{s}
=|T^{\uparrow}|^{2}\varsigma_{1+}+|T^{\downarrow}|^{2}\varsigma_{2+}-Z_{+}\sigma_{z}
+d_{1}\varpi_{1+}+d_{2}\varpi_{2+},
\end{equation}
where
$Z_{\pm}=|R|^{2}\pm(T^{\uparrow *}T^{\downarrow}+T^{\downarrow *}T^{\uparrow})
\cos^{2}\frac{\theta}{2}\sin^{2}\frac{\theta}{2}$;
$d_{1}=T^{\uparrow *}R-R^{*}T^{\downarrow}$;
$d_{2}=T^{\downarrow *}R-R^{*}T^{\uparrow}$, and
\begin{eqnarray}
\varsigma_{1\pm}&=&\left(
\begin{array}{cc}
\cos^{4}\theta/2 & 0 \\
0 & \pm\sin^{4}\theta/2
\end{array}
\right);\\
\varsigma _{2\pm}&=&\left(
\begin{array}{cc}
\sin^{4}\theta/2 & 0 \\
0 & \pm\cos^{4}\theta/2
\end{array}
\right);
\end{eqnarray}
\begin{eqnarray}
\varpi_{1\pm}&=&\left(
\begin{array}{cc}
0 & \cos^{2}\theta/2 e^{-i\varphi} \\
\pm\sin^{2}\theta/2 e^{i\varphi} & 0
\end{array}
\right);\\
\varpi_{2\pm}&=&\left(
\begin{array}{cc}
0 & \sin^{2}\theta/2 e^{-i\varphi} \\
\pm\cos^{2}\theta/2 e^{i\varphi} & 0
\end{array}
\right).
\end{eqnarray}
Using above equations we can give straightforwardly Eqs. (\ref{Pgs}) and (\ref{Pjs}).

\end{document}